\documentstyle[preprint,aps]{revtex}
\begin{document}
\draft
\title{
DOF Phase Separation of the Lennard-Jones fcc(111) Surface
}
\author{C. S. Jayanthi$^{\rm (1,2)}$, Franck Celestini$^{\rm (3,4,5)}$,
        Furio Ercolessi $^{\rm (3,4)}$, and
        Erio Tosatti $^{\rm (1,3,4)}$}
\address{
$^{\rm (1)}$ The Abdus Salam
International Centre for Theoretical Physics (ICTP),
I-34014 Trieste, Italy}
\address{
$^{\rm (2)}$ Department of Physics, University of Louisville,
Louisville, KY 40292, USA}
\address{
$^{\rm (3)}$ International School for Advanced Studies (SISSA-ISAS),
I-34014 Trieste, Italy}
\address{
$^{\rm (4)}$ Istituto Nazionale di Fisica della Materia (INFM), Unit\`a
Trieste-SISSA}
\address{
$^{\rm (5)}$ Laboratoire MATOP, Universite D'Aix-Marseille III,
F-13397 Marseille, France}


\maketitle
\begin{abstract}
Recent lattice model calculations have suggested that a full-layered 
crystal surface may undergo,
under canonical (particle-conserving) conditions,
a preroughening-driven two-dimensional phase separation into two 
disordered flat (DOF) regions, of opposite order parameter.
We have carried out extensive classical molecular dynamics (MD)
simulations of the Lennard-Jones fcc(111) surface, to check
whether these predictions are relevant or not for a realistic 
continuous system.
Very long simulation times, a grid of temperatures from $(2/3)T_m$
to $T_m$, and unusually large system sizes are employed to
ensure full equilibrium and good statistics.
By examining layer-by-layer occupancies, height fluctuations, sublattice 
order parameter and X-ray structure factors, we find a clear 
anomaly at $\sim 0.83\,T_m$. The anomaly is distinct from roughening
(whose incipiency is also detected at $\sim 0.94\,T_m$), and is
seen to be consistent with the preroughening plus phase separation 
scenario. 
\end{abstract}
\vspace{2cm}

\noindent
\begin{center}{\bf SISSA Ref. 56/99/CM/SS (revised April 2000)}.
\end{center}

\newpage
\tightenlines
The surfaces of a rare-gas solid such as Ar are reasonably well modeled
by those of a (truncated) Lennard-Jones (LJ) fcc solid.
The behavior of the LJ surfaces as a function of temperature, particularly
in the vicinity of the melting point $T_m$, has been the subject of 
a large number of studies, mostly by classical Molecular Dynamics (MD)
\cite{LJsurf}.
Based on these studies, the general consensus until recently was that
the LJ surfaces begin to disorder, with increasing temperature $T$,
in a very gradual manner.
As $T\gtrsim  (2/3)\,T_m$,  surface anharmonicities \cite{JTP85}
and defects \cite{LJsurf} first build up.
In particular, there is a progressive growth in the number of surface
adatoms/vacancies in the region $T\sim 0.7 \div 0.8 \,T_m$ and above.
Their presence undermines the surface crystalline state \cite{Stoltze89}.
Eventually, a microscopic quasi-liquid surface film is formed,
which leads to surface melting as $T_m$ is approached \cite{Dash89}.
En route to surface melting, a surface roughening transition at
$T_{\rm R} < T_m$ should also appear \cite{LT86}, 
owing to a step free energy softening
in presence of the quasi-liquid film.
On Ar(111), for example, it is known that 
$T_{\rm R} \simeq 0.94\,T_m$ \cite{Larher91}.

Subsequently, however, experimental evidence has appeared
\cite{Youn90,Day93,Youn93,Rieutord97}, followed by theoretical work
\cite{denNijs91,Weichman95,PST95,Weichman98,Prestip99,Jagla99} which upsets 
this gradual picture, and suggests that surface disordering occurs 
instead through a singularity.
In connection with statistical mechanics models introduced by
Den Nijs in the 80's \cite{Rommelse87},
this singularity is probably best understood as ``preroughening'' (PR).
At PR, the originally ordered flat surface turns into a 
disordered flat (DOF) state, where surface steps proliferate, 
albeit with strict up-down alternation.
In this way surface flatness is preserved while still gaining entropy
from disorder, in particular from step meandering \cite{denNijs91}.
Due to the resulting checkerboard texture of steps, a DOF surface phase
exhibits, at least in the lattice models, a striking {\em half-integer}
occupancy in the topmost layer.

More recently, grand canonical Monte Carlo simulations \cite{GCMC1,GCMC2}
have confirmed the occurrence of a PR phase transition, and of an associated
coverage jump at the LJ(111) surface, in the neighborhood of $0.83\,T_m$.   

The questions we address in this note are the following.
Can we first of all detect preroughening and roughening in a standard
canonical (that is, particle-conserving) realistic surface MD simulation?
And, if the PR scenario is indeed correct for the LJ surfaces, 
how do PR and the appearance of a DOF phase exactly manifest themselves?  
Finally, why did such an important 
singularity go unnoticed in so many previous, 
good-quality MD simulations?

We surmised recently \cite{PT98} that the answer might be 
that in MD, the singularity is masked by particle conservation,
turning a sharp critical onset into a subtler, Ising-like
phase separation. A Monte Carlo study of a lattice RSOS (restricted
solid-on-solid) model did indeed show that under canonical, 
particle-conserving conditions, an initially full surface monolayer 
spontaneously phase separates above $T_{\rm PR}$ into two DOF regions, 
each of which has essentially population 1/2 in the top layer \cite{PT98}.
However, lattice models may be oversimplified, and a more realistic
study of this question is strongly needed.

Here, we describe new extensive canonical MD simulations of the 
LJ fcc(111) surface,
specifically aimed at understanding whether the DOF phase separation is
real or not.  Anticipating our conclusions, we shall find that the 
answer is affirmative. However we also find that the evidence can 
only be obtained by employing large sizes, long simulation times, and 
size scaling, usually not considered in previous work. A corollary is
that these kinds of precautions, aimed at detecting possible DOF surface
phase separation, appear mandatory for future studies
of warm surfaces using canonical MD simulation.

We conducted all our simulations in a slab geometry, with $N_L$ fully mobile
layers of $M$ particles/layer and $(x,y)$ periodic boundary conditions.
Three rigid fcc(111) layers are added at the bottom of the slab.
To account for thermal expansion, the $(x,y)$ simulation cell size
was expanded with temperature according to an expansion coefficient
extracted from an independent set of bulk simulations.
Since we are aiming at describing a solid-vapor interface at 
full thermodynamic equilibrium, our simulation allowed an approximately
equal volume of LJ gas above the solid surface. The gas particles 
were contained and backscattered by a perfectly reflecting wall 
placed at a distance $N_V$ ``layers'' in the vacuum above the last 
crystal layer.

We studied in detail three samples:
{\em (i)} SF (small full) where $M=120$ particles/layer, 
	$N_L =13$, $N_V \simeq 10$;
{\em (ii)} LF (large full) where $M=504$, $N_L=25$, $N_V \simeq 15$;
{\em (iii)} LH (large half) where $M=504$, $N_L=24.5$, $N_V \simeq 15$.
The interparticle interaction was 12-6 LJ, truncated at $2.5\,\sigma$.
The bulk melting temperature for this system was determined
to be $k_B T_m \simeq 0.7\,\epsilon$, by observing that at
this temperature a relatively large number ($\approx 4$) of melted 
layers were stable while the rate of growth of the liquid film
was maximal (full melting of the slab being prevented by
the three bottom rigid layers).

A standard fifth-order predictor-corrector algorithm was used 
for the time integration of Newton's equation, with a time step of
0.01 LJ units, amounting to $2.15\times 10^{-14}\,\rm s$
in the case of Ar ($\epsilon/k_B = 120\,\rm K$, $\sigma = 3.40\,\rm\AA$).
Temperature control based on velocity rescaling was used to heat or cool
the system. However, all the equilibration and data collection runs
were done in constant energy, microcanonical conditions.

In each case, the standard procedure adopted was to equilibrate the
system for $10^5$ steps, at each temperature $T$ chosen in the interval
$\sim 0.7\,T_m \div T_m$.
During this time (amounting to about 2 ns for Ar) some surface atoms
evaporated and others recondensed.
However, their number (less than 10 in all cases) was negligible 
in comparison with one monolayer, so that the solid underneath 
was basically running under particle conserving conditions.
Since surface evolution in these conditions is strictly determined by
diffusion, long equilibration times are dictated by the requirement that
the lateral diffusion length of a surface particle should be roughly
comparable to half the $(x,y)$ cell size.
With a diffusion coefficient of the order of 
$0.2\times 10^{-4}\,\rm cm^2/s$ for Ar at $T_m$ \cite{ardiff},
2 ns are sufficient to ensure equilibration by diffusion.
Following equilibration, we ran a further 
$5\times 10^4 \div 10^5$ steps for a good statistical average of
thermodynamical quantities, which we now proceed to discuss.

{\em (i) Height fluctuations.}
The surface height fluctuations were obtained by calculating
$$\delta h^2 = 
\left\langle
\frac{1}{N_S} \sum_i (h_i - \bar{h} )^2
\right\rangle
$$
where $h_i$ denotes the $z$-coordinates of all {\em surface particles}
$i$ whose identity and total number $N_S$ are defined at each
given configuration, along with the average height
$\bar{h} = (1/N_S) \sum_i h_i$.
Configurations were chosen at regular time intervals (one every 250 steps
for SF, every 500 steps for LF, LH),
and $\langle\ldots\rangle$ denotes average over configurations.
Surface particles were identified as those not totally or even partially 
shadowed by other particles, when viewing the surface from the
gas and parallel to the $z$ axis, each particle 
being represented as a sphere of atomic radius $r_\circ = 0.9\,\sigma$.

A flat, defect free vibrating crystal surface should be characterized by
$\delta h^2 \sim d^2$, where $d$ is the interlayer spacing.
Proliferation of surface adatoms/vacancies should give rise to
an increase of $\delta h^2$, which for increasing size should tend 
to a finite value, so long as the surface is flat, {\em i.e.}\ non-rough.
Surface roughening should be further signalled by a  
size-dependent $\delta h^2$ increase, scaling with size as $\log M$.
Below roughening and neglecting vibrations,
an ideal DOF surface, with its half-occupied topmost layer,
should be characterized by $\delta h^2 = (1/4) d^2$.
Conversely, a full-layered surface, phase-separated into
two DOF domains with a height difference of $d$ between them,
would instead exhibit a larger $\delta h^2 \simeq (1/2) d^2$.

Fig.\ \ref{fig:height} shows the behavior obtained for $\delta h^2$.
There is a clear and sudden change in the full-surface
behavior near a breakdown temperature $\sim 0.83\,T_m$.
The height fluctuation $\delta h^2$ grows fast for both SF and LF 
below this temperature; above, it levels off to a value close to that of LH.
Moreover, comparison of SF and LF indicates a stronger size-dependence
of $\delta h^2$ at the breakdown temperature.
Both features, in accordance with previous discussions based on
the FCSOS model \cite{PT98},
strongly suggest that {\em preroughening} is taking place
at the breakdown temperature, $T_{\rm PR}\simeq 0.83\,T_m$.
A strong size-dependence of $\delta h^2$
reappears again at higher temperature, compatible with an
estimated roughening temperature, $T_{\rm R} \simeq 0.94\,T_m$.
Both values are in excellent agreement with the experimentally 
established values of $T_{\rm PR}/T_m \simeq 69/84$, obtained
from reentrant layering of Ar(111)/graphite \cite{Youn93}, and of
$T_{\rm R}/T_m \simeq 80/84$, for roughening \cite{Larher91}.
This is, to our knowledge, the first determination of these temperatures
ever obtained by direct molecular dynamics simulation.

Between $T_{\rm PR}$ and $T_{\rm R}$, the canonical SF and LF surfaces 
should really consist, according to the lattice model, of domains 
made of two kinds of phase-separated DOF regions.
The height fluctuations do not show strong evidence for that,
because $\delta h^2$ remains close to $(1/4) d^2$ instead of $(1/2) d^2$. 
We will return
to this discrepancy later, when discussing evidence for surface melting.

Next we calculated the layer occupancies, shown in fig.\ \ref{fig:occup},
as a crucial indicator of phase separation \cite{PT98}.
Due to phase separation, we would expect that
(neglecting vacancies in the second and deeper layers)
the concentration of vacancies in the first layer, $n_v$, and 
of adatoms one layer above, $n_a$, should be
$n_v = N_v/M = (1-\sigma)/2$, 
$n_a = N_a/M = \sigma/2$,
with $\sigma$ the ``Ising magnetization'' of this problem.
In particular we expect $\sigma=1/2$ for SF, LF, and
$\sigma=0$ for LH, whence 
$n_v = 1/2$, $n_a = 0$ for a single DOF phase in LH, but
$n_v = n_a = 1/4$ for two phase-separated DOF's in SF, LF.
Hence at the onset of DOF phase separation these concentrations,
ordinarily growing with temperature, should stabilize around 1/4
and roughly stop growing until roughening.

We analyse first the full-layer results, SF and LF. 
Fig.\ \ref{fig:occup} shows the evolution of the vacancy and adatom
concentration with temperature. Beginning with initially small
concentrations, the growth rate with temperature 
has a kink-like feature near $0.83\,T_m$.  Comparing SF an LF, 
the kink becomes more pronounced with increasing size, particularly
for adatoms. Above the kink, there is a visible tendency to 
stabilize concentrations at a plateau value which,
for both $n_a$ and $n_v$, lies between 0.2 and 0.25, only slightly smaller
than the expected phase separation value of 1/4. A possible reason 
why $n_a$ and $n_v$ might tend to fall slightly below 1/4  could
possibly be traced to a boundary effect
between one DOF domain and the other.
At the boundary, two {\em parallel} steps occur near one another,
and parallel step repulsion \cite{PST95} could be expected to
push them apart, thus reducing in the neighborhood
both $n_a$ and $n_v$ below $1/4$.
At much higher temperatures, $n_v$ nicely tends to $1/4$, while $n_a$
grows again in the neighborhood of $0.94\,T_m$, where we believe
that roughening is taking place. We conclude that the overall 
behavior of adatom/vacancy concentration in the full-layer
simulations provides a 
clear evidence for DOF phase separation above the break at $0.83\,T_m$.

Next, we consider the half-layer system LH which, by contrast, 
shows (Fig.\ \ref{fig:occup})  $n_a$ and $n_v$ to remain
close to $0$ and to $0.5$, in full equilibrium,
across the whole temperature range, with no particular features throughout.
While the presence of two built-in antiparallel steps represents
a trivial separation of ordered flat phases at low temperatures,  we 
can conclude from the lack of further evolution of adatom/vacancy
concentrations that only the full-layered surfaces show
a tendency to spontaneously phase separate above $T_{\rm PR}$, fully
consistent with expectations.

It should be possible to identify some of these features by a direct
examination of the physical structure of the surfaces generated by the
simulation. Fig.\ \ref{fig:snapshots} shows snapshots of the full-
and half-covered surface at the end of runs below and above $0.83\,T_m$.
In spite of the fuzziness necessarily present in such snapshots,
the following features can be recognized:
{\it a)} the low-$T$ full surface (top left) is relatively crystalline,
although with a large number of adatoms and vacancies (this is about
twice the average value in fig.\ \ref{fig:occup});
{\it b)} the high-$T$ full surface (top right) is now spread on two
layers (white and grey), indicating phase separation;
{\it c)} the low-$T$ half-covered surface (bottom left) is also spread
on two layers (gray and black), again indicating phase separation;
{\it d)} the high-$T$ half-covered surface (bottom right) is very 
disordered, indicating surface melting.

These results, providing a vivid picture of what the actual surface
instantaneously looks like, are in support of the occurrence of PR
between the two temperatures.
Moreover, they indicate melting of the outermost layer above PR,
a result predicted theoretically \cite{Jagla99} and independently
obtained by Grand Canonical Monte Carlo simulations \cite{GCMC2}.

As a final piece of evidence, we have calculated the
sublattice, or DOF, order parameter, defined as
$$
P = \left\langle
\frac{1}{N_S} \sum_i {\rm e}^{i\pi h_i/d} .
\right\rangle
$$
The order parameter is finite for an ordered flat surface; it should
drop to zero precisely at $T_{\rm PR}$. Due to a lack of adatom-vacancy
symmetry, it should be again finite, but substantially
smaller, between $T_{\rm PR}$ and $T_{\rm R}$. The drop
should be either continuous or abrupt, according to whether PR was
critical or first order.
This quantity has the additional merit of corresponding 
roughly to the X-ray antiphase scattering amplitude, to be expected 
for such a surface \cite{mazzeo}.

As seen in fig.\ \ref{fig:orderpar}, $P$ has a drop, for both SF and LF,
more visible for LF, near $0.83\,T_m$. Comparison of the two 
indicates a weak but nonzero size dependence, suggesting 
a probable weak first order nature for PR in this surface.

Conversely, the half-layer surface LH exhibits simply 
a monotonic increase of $P$,
roughly up to $T_{\rm R}$, where results for all three systems,
SF, LF, LH eventually merge. When interpreting results
for LH below $T_{\rm PR}$ and for SF, LF above $T_{\rm PR}$, we
should keep in mind that the order parameter average is not 
particularly meaningful for a phase-separated surface.  
We expect that, as a rough approximation, 
the true grand-canonical surface would show values of $P$ similar to LF
for $T<T_{\rm PR}$, with a jump to those of LH for $T>T_{\rm PR}$
(dashed line).

Finally, we wish to return briefly to the open problem of anomalously
small height fluctuations found earlier in the DOF separated
region of systems SF and LF. It seems possible that an explanation
could be found if the top layer did develop a liquid-like 
character above 0.83 $T_m$. In conjunction
with relatively small domain sizes, high surface diffusion 
would lead to a globally much more homogeneous, and thus flatter, 
surface than could be expected on the bare, rigid phase separation picture. 
The high mobility scenario would apply in particular if, as 
has been repeatedly suggested, the first
surface monolayer, or fraction of monolayer, is in reality
solid below $0.83\,T_m$ but melted above \cite{phillips,GCMC2}.
We qualitatively inspected the  diffusivity of our surface
atoms and found it indeed to be higher for LH. A fully quantitative
investigation, reported separately \cite{GCMC2}, 
confirmed the existence
of a jump of the top layer lateral diffusion coefficient by more 
than a factor 2 between LF and LH at $T<T_{\rm PR}$. That result does suggest
that the first layer is premelting at the same time as it is 
preroughening, and indirectly also provides an explanation for
the global homogeneity of the DOF phase separated outer surface layers.
 
In summary, we have found that a canonical, particle conserving 
MD simulation for an fcc(111) LJ surface contains clear indications for
a) preroughening at $T_{\rm PR} = 0.83\,T_m$; b) roughening at 
$T_{\rm R} = 0.94\,T_m$;  c) a DOF phase separation between $T_{\rm PR}$ 
and $T_{\rm R}$. The latter result is of conceptual relevance, since
it shows that the continuous, off-lattice degrees of freedom of
the real system do not destroy, with some qualifications, the 
essence of DOF physics, previously established only in much less
realistic lattice models.

Direct experimental observation of DOF phase separation will not be 
possible on the solid rare gas surfaces, where the very high evaporation
rates immediately establish grand canonical equilibrium. In this sense
our study is academic for that system, for which we developed
separately a grand canonical Monte Carlo approach \cite{GCMC1,GCMC2}. 
On the other hand, the DOF phase separation 
described here could in real life be realized on metals, under
conditions where evaporation is irrelevant.

Independently of experimental accessibility, there is an important 
methodological message to be drawn from the present work, which is 
very pertinent to the simulation community: surfaces which are likely 
to undergo a DOF transition should as a rule be simulated grand 
canonically. Canonical molecular dynamics simulation in particular
is very dangerous, as it may inadvertently describe an unphysical 
state of the surface in the whole temperature interval between
preroughening and roughening.

We are grateful to Santi Prestipino for enlightening discussions.
Work at SISSA by  F.\ C.\ was under European Commission sponsorship,
contract ERBCHBGCT940636. We acknowledge support from INFM PRA LOTUS,
and by MURST. C.\ S.\ J.\ is grateful to the ICTP for hospitality.



\begin{figure}
\caption{
Height fluctuations $\delta h^2$ as a function of temperature
for the small full (SF, black squares), large full (LF, black circles)
and large half-occupied (LH, white circles) sample.
Note the gap between LF and LH at low temperature, closing up 
around $0.83\,T_m$.
Other features are discussed in the text.
}
\label{fig:height}
\end{figure}

\begin{figure}
\caption{
Layer occupancies as a function of temperature.
Symbols are as in fig.\ \ref{fig:height}.
Note how the vacancy and adatom and concentrations of the 
full-layer samples (SF, LF)
tend to stabilize in the proximity of 1/4 between about
$0.83\,T_m$ and $0.94\,T_m$, consistent with the presence of
two separated DOF phases in this region.
}
\label{fig:occup}
\end{figure}

\begin{figure}
\caption{
Top view of instantaneous snapshots of the three outer layers
for systems with initial full- (top) and half-layer (bottom) occupation,
at $0.76\,T_m$ (left) and $0.85\,T_m$ (right).
Atoms have been colored in black, grey or white according to the
layer they belong to, in turn determined from their $z$ coordinate
and the density profile along $z$ of each sample.
}
\label{fig:snapshots}
\end{figure}

\begin{figure}
\caption{
DOF parity order parameter as a function of temperature.
Symbols are as in fig.\ \ref{fig:height}.
Note the drop around $0.83\,T_m$ for the full-layer samples (SF, LF),
indicating tendency toward a two-level structure.
A hypothetical grand canonical system would approximately follow the 
dashed line, drastically changing the layer population at $T_{\rm PR}$
in correspondence with a large decrease of the order parameter.
}
\label{fig:orderpar}
\end{figure}
\end{document}